\documentclass[12pt]{article}
\usepackage{a4wide}
\usepackage{graphicx}
\usepackage{amssymb}
\usepackage{hyper}
\newcommand{\be}{\begin{equation}}
\newcommand{\ee}{\end{equation}}
\newcommand{\ba}{\begin{eqnarray}}
\newcommand{\ea}{\end{eqnarray}}
 
\newcommand{\pdir}{p\kern -5.2pt\raise 0.2ex\hbox {/}}
\newcommand{\vdir}{v\kern -5.75pt\raise 0.15ex\hbox {/}}
\newcommand{\kdir}{k\kern -5.75pt\raise 0.15ex\hbox {/}}
\newcommand{\epsdir}{\epsilon\kern -5.0pt\raise 0.15ex\hbox {/}}
\newcommand{\bvdir}{\bar{v}\kern -5.75pt\raise 0.15ex\hbox {/}}
\newcommand{\Ddir}{D\kern -7.75pt\raise 0.20ex\hbox {/}}
\newcommand{\ldir}{l\kern -5.0pt\raise 0.2ex\hbox{/}}
\newcommand{\Tr}[1]{\mathrm{Tr}\left[#1\right]}
\begin{document}

\begin{titlepage}
\begin{flushright}
LU TP 10-16\\
arXiv:1006.1197 [hep-ph]\\
revised October 2010
\end{flushright}
\vfill
\begin{center}
{\Large\bf Hard Pion Chiral Perturbation Theory for $B\to\pi$  and $D\to\pi$
Formfactors}
\vfill
{\bf Johan Bijnens and Ilaria Jemos}\\[0.3cm]
{Department of Astronomy and Theoretical Physics, Lund University,\\
S\"olvegatan 14A, SE 223-62 Lund, Sweden}
\end{center}
\vfill
\begin{abstract}
We use one-loop Heavy Meson Chiral Perturbation Theory (HMChPT) 
as well as a relativistic formulation to
calculate the chiral logarithms $m^2_\pi\log{\left(m^2_\pi/\mu^2\right)}$
contributing to the formfactors of the semileptonic $B\rightarrow \pi$ decays
at momentum transfer $q^2$ away from $q^2_\mathrm{max}=(m_B-m_\pi)^2$.
We give arguments why this chiral behavior is reliable even in the energy
regime with hard or fast pions. These results can be used to extrapolate
the formfactors calculated on the lattice to lower light meson
masses.
\end{abstract}
\vfill
\end{titlepage}

\section{Introduction}

The study of the formfactors of semileptonic $B\rightarrow\pi$
and $D\rightarrow\pi$ decays has
become a task of primary importance for the determination of the KM matrix
elements $\left|V_{ub}\right|$ and $\left|V_{cd}\right|$ respectively.
Unfortunately this is a rather difficult mission. The kinematically
accessible region is large, the physical pictures emerging at the two
extremities of the $q^2$ range are quite different, requiring different
approximation methods  and this makes 
calculations directly from QCD hard.

Nevertheless a lot of effort has been put into studying this decay,
both in experiment and in theory. 
The $q^2$ spectrum has been measured by different
collaborations (CLEO \cite{:2008yi,Adam:2007pv}, Belle
\cite{Hokuue:2006nr}, Babar \cite{Aubert:2006px} )
The QCD based theoretical calculations are either on QCD light-cone sum rules 
(LCSR), which provide reliable determinations at small $q^2$ \cite{Duplancic:2008ix},
or on lattice simulations \cite{Gamiz:2008iv}.
In particular lattice QCD allows to solve the non perturbative QCD effects
numerically, but it
is at present limited in the light quark masses that can be reached. Thus a
final extrapolation in the light quark masses is needed. 

At low energies Chiral Perturbation Theory (ChPT) \cite{GL1,GL2} provides a
way to do this extrapolation on a theoretically sound basis.
For processes with all pions soft this works fine and was extended to
include heavy mesons in  \cite{Wise:1992hn,Burdman:1992gh}.
The $B$ and $B^*$ mesons were there included using a heavy quark like formalism
known as Heavy Meson ChPT (HMChPT).
This has been used to extrapolate the behaviour of the form factors
near the endpoint, $q^2_\mathrm{max}=\left(m_B-m_\pi\right)^2$, where
the pions are soft \cite{Falk:1993fr,Becirevic:2003ad}.
But a description of the light quark mass dependence of the form
factors in the entire range of energy is still missing. Therefore we lack
extrapolation formulas in the region away from maximum momentum transfer.

A similar problem exists in the case of $K_{\ell 3}$ decay. Here two-flavour
$(SU(2))$ ChPT provides a well defined scheme for the
calculation of the formfactors near $q^2_\mathrm{max}=\left(m_K-m_\pi\right)^2$.
At other values of $q^2$ including $q^2\simeq0$,
the (two-flavour) power counting scheme breaks down due to the presence
of a large momentum pion in the final state. 
However, the authors of \cite{Flynn:2008tg} argued that also
in this latter case the coefficient of the chiral logarithm
$m^2_\pi \log{m^2_\pi}$ is calculable and thus it can be used for the
extrapolations on the lattice at $q^2$ away from $q^2_\mathrm{max}$.
In \cite{Bijnens:2009yr} the argument was clarified and extended to the case
of the $K\rightarrow \pi\pi$ decays. It was also argued there that this was
a much more general circumstance. 

The aim of this paper is to perform the same calculations for
heavy meson semileptonic decays. These results can then be used to perform the
extrapolation to light quark masses of lattice results also for values of $q^2$
away from the end-point $q^2_\mathrm{max}$. The arguments as presented
in \cite{Bijnens:2009yr} show that also in this case the coefficient of
the logarithm should be calculable as discussed in Sect.~\ref{argument}.
The discussion implies that both the HMChPT formalism or a relativistic one
can be used. We have
performed the calculations in both formalisms as a consistency check and have
also reproduced the known results for the masses, decay constants and
formfactors at $q^2_\mathrm{max}$ in both.

The paper is structured as follow. After a short description of HMChPT  in
Sect.~\ref{HMChPT}, we introduce in Sect.~\ref{relth} the relativistic
Lagrangian that we used as a consistency check, since
the off-shell behaviour in both formalisms is rather different.
In Sect.~\ref{Bl3:formalism}  we define the formfactors involved, 
and how to include the weak current in the Lagrangians of both formalisms.
Sect.~\ref{argument} gives the arguments why this procedure should produce the
correct nonanalytic behaviour in the light quark masses where they are
different from \cite{Flynn:2008tg,Bijnens:2009yr}.
Finally the results for the coefficients are shown in
Sect.~\ref{ChLogs} where we provide also some checks of the validity
of our assumptions. The appendix gives some results for the needed expansions
of the loop integrals.

Throughout the paper we focus on $B\rightarrow\pi\ell\nu_\ell$ decay,
but the same procedure and calculations go through also in the $D$
semileptonic decays. All formulas are applicable to both cases.
We are extending this work to the three flavour case
as well as to other vector formfactors like $B\rightarrow D$ \cite{nextpaper}.

\section{Heavy Meson Chiral Perturbation Theory}\label{HMChPT}

In this section we review the main features of HMChPT
\cite{Wise:1992hn,Burdman:1992gh}, see also the lectures by Wise
 \cite{Wise:1993wa} and the book \cite{Heavyquarkbook}.
Chiral Lagrangians can be used to describe the interactions of light mesons,
as pions and kaons, with hadrons containing a heavy quark. HMChPT makes use of
spontaneously broken $SU(N_f)\times SU(N_f)$ chiral symmetry on the light
quarks, and
spin-flavour symmetry on the heavy quarks. This formulation lets us study
chiral symmetry breaking effects in a chiral-loop expansion by simultaneously
performing an expansion in powers of the inverse of the heavy meson mass.
 
In this paper we deal only with two-flavour ChPT \cite{GL1}  but the theory can
be easily extended in the case of three flavours \cite{GL2},
thus including kaons in the description. 
The notation is the same as in \cite{Bijnens:1999sh}.
The lowest order Lagrangian describing the strong interactions of the light
mesons is
\be
\label{pilagrangian}
\mathcal{L}_{\pi \pi}^{(2)}= \frac{F^2}{4}  
\left( \langle u_{\mu} u^{\mu} \rangle   + \langle \chi_{+}\rangle \right),
\ee
with
\ba
 u_{\mu} &=& i\{  u^{\dag}( \partial_{\mu} - i r_{\mu}   )u
 -u ( \partial_{\mu}   -i l_{\mu}    ) u^{\dag}   \}\,,
\nonumber\\
\chi_{\pm} &=& u^{\dag} \chi u^{\dag} \pm u \chi^{\dag} u\,,
\nonumber\\
u &=& \exp\left(\frac{i}{\sqrt{2}F} \phi \right)\,,
\nonumber\\
 \chi &=& 2B (s+ip) ,\nonumber \\
\phi &=&
 \left( \begin{array}{cc}
\frac{1}{\sqrt{2}}\pi^0  &  \pi^+  \\
 \pi^- & -\frac{1}{\sqrt{2}}\pi^0 \\
 \end{array} \right)\,.
\ea
The fields $s$, $p$, $l_\mu=v_\mu-a_\mu$ and
$r_\mu=v_\mu+a_\mu$ are the standard external scalar, pseudoscalar, left- and
right- handed vector fields introduced by Gasser and Leutwyler \cite{GL1,GL2}.

The field $u$ and $u_\mu$ transform under a chiral transformation $g_L\times g_R
\in SU(2)_L\times SU(2)_R$ as
\ba
\label{trasfrules}
u \longrightarrow g_R u h^\dagger = h u g_L^\dagger,\qquad
u_\mu\longrightarrow h u_\mu h^\dagger.
\ea
In (\ref{trasfrules}) $h$ depends on $u$, $g_L$ and $g_R$ and
 is the so called compensator field.
The notation $\langle X\rangle$ stands for trace over up and down quark
indices and all matrices are $2\times2$ matrices. 

We now begin with a brief synopsis of the formalism of HMChPT for the
two-flavour case. The three flavour case was the original formulation
\cite{Wise:1992hn,Burdman:1992gh}. In the limit $m_b\rightarrow\infty$,
the pseudoscalar $B$ and the vector $B^*$ mesons are degenerate. In the
following we neglect the mass splitting $\Delta=m_{B^*}-m_B$.
To implement the heavy quark symmetries it is convenient to assemble
them into a single field
\ba
\label{heavyfield}
H^a(v) &=& {1 +  \vdir \over 2} \left[ B^{\ast a}_\mu (v)\gamma^\mu - B^a
  (v)\gamma_5\right],
\ea
where $v$ is the fixed four-velocity of the heavy meson, $a$ is a flavour
index corresponding to the light quark in the $B$ meson.
$B^1 = B^+$, $B^2=B^0$ and similarly for the vector mesons $B^*_\mu$.
In (\ref{heavyfield}) the operator
$(1+\vdir)/2$ projects out the particle component of the heavy meson only.
The conjugate field is defined as 
$ \overline H_a(v) = \gamma_0 H_a^\dagger (v) \gamma_0$. We assume the field
 $H^a(v)$ to
 transform under the chiral transformation $g_L\times g_R
\in SU(2)_L\times SU(2)_R$ as
\be
H_a(v) \longrightarrow h_{ab}H_b(v)\,,
\ee
so we introduce the covariant derivative as 
\ba
D^\mu_{ab}H_b(v) = \delta_{ab} \partial^\mu H_b(v)  +\Gamma^\mu_{ab}H_b(v),
\ea
where 
$\Gamma^\mu_{ab} =
\frac{1}{2}\left[u^\dagger\left(\partial_\mu-ir_\mu\right)u+u\left(\partial_\mu-il_\mu\right)u^\dagger\right]_{ab}$, 
 and the indices $a$, $b$ run over the
light quark flavours. 
Finally, the 
Lagrangian for the heavy-light mesons in the static heavy quark limit reads 
\ba 
\label{heavylag}
\mathcal{L}_\mathrm{heavy}=-i\,\Tr{\overline{H}_a  v \cdot D_{ab}H_b}+g\,
\Tr{\overline{H}_a u_{ab}^\mu H_b \gamma_\mu \gamma_5},
\ea
where $g$ is the coupling of the heavy meson doublet to the Goldstone boson
and the traces, $\mathrm{Tr}$, are over spin indices,
the $\gamma$-matrix indices.
The Lagrangian (\ref{heavylag}) satisfies chiral symmetry and heavy
quark spin flavour
symmetry.

As a final remark of this section we stress that, in general,
the use of HMChPT is only valid as long as the interacting pion is soft,
i.e. if it has momentum much smaller than the scale of spontaneous chiral
symmetry breaking ($\Lambda_\mathrm{ChSB}\simeq 1~\mathrm{ GeV}$). 
In fact, only in this regime the usual ChPT is well defined. For
the semileptonic decays of heavy mesons this range of energy covers just a
small fraction of the Dalitz plot. In Sect.~\ref{argument} we will give an
argument why the predictions on the coefficients of the logarithms appearing
in the final amplitudes are reliable even outside the range of applicability
of HMChPT. 

\section{Relativistic Theory}
\label{relth}

When $q^2 \ne q_\mathrm{max}$ it is possible that in the loops appear very
off-shell $B$ and $B^*$ mesons. This in principle changes the non analyticities
in the light masses of the loop functions and thus it might affect
the coefficients of $m_\pi^2\log{m^2_\pi/\mu^2}$. It could be that different
treatments of the off-shell behaviour gave rise to different nonanalyticities.
Sect.~\ref{argument} argues that this should not be the case.
In order to test this, we are not only calculating using HMChPT
but also in a relativistic formulation. We also add some redundant higher order
terms as an additional check.

For this scope, we construct a relativistic
Lagrangian that respects the
spin-flavour symmetries of HMChPT. It is built up starting from $B^a$
and $B^{*a}_\mu$ fields, but now in the relativistic form,
and we treat them as column-vectors in the light-flavour index $a$.
\ba
\label{rellagkin}
\mathcal{L}_\mathrm{kin}&=&
\nabla^\mu B^\dagger \nabla_\mu B -m_BB^\dagger B
-\frac{1}{2}B^{*\dagger}_{\mu \nu}B^{*\mu \nu}
+m_BB^{*\dagger}_{\mu}B^{*\mu},\\
\mathcal{L}_\mathrm{int} &=&g M_0\left(B^\dagger u^\mu
B^{*}_\mu+B^{*\dagger}_\mu u^\mu
B\right)\nonumber\\
&+&\frac{g}{2}\epsilon^{\mu\nu\alpha\beta}\left(-B^{*\dagger}_\mu
u_\alpha\nabla_\mu B^*_\beta+\nabla_\mu B^{*\dagger}_\nu
u_\alpha B^*_\beta\right),\label{rellagint}
\ea
with
$
B^*_{\mu \nu}=\nabla_\mu B^*_\nu-\nabla_\nu B^*_\mu$, and
$\nabla_\mu =
\partial_\mu+\Gamma_\mu$.
The constant $g$ of (\ref{rellagint}) is the same in
(\ref{heavylag}), $M_0$ is the mass of the $B$ meson in the chiral limit. In
(\ref{rellagkin}) and (\ref{rellagint})
we have suppressed flavour indices $a$, $b$ for simplicity. The fields $B$
and $B^*$ transform under chiral transformations as $B\to hB$.
The two terms of $\mathcal{L}_\mathrm{int}$ in (\ref{rellagint}) contain
 the vertices $B B^{*} \pi$ and
$B^{*}B^{*}\pi$. No
interaction of the kind $B B \pi$ appears because it is forbidden by parity
conservation. 

{}From $\mathcal{L}_\mathrm{kin}$ in (\ref{rellagkin}) we find the
propagators of the $B$ and $B^{*}$ meson respectively:
\be
\frac{i}{p^2-m_B^2}, \qquad \frac {-i\left(g_{\mu \nu}-\frac{p_\mu p_\nu}{m^2_B}\right)}{p^2-m^2_B}.
\ee
This is to be contrasted with the propagator $1/v\cdot p$ in the HMChPT
formalism showing the different off-shell behaviour.

\section{$B\to\pi$ formfactors: formalism}
\label{Bl3:formalism}

In this section we review the semileptonic decay formalism.
The hadronic current for pseudoscalar to pseudoscalar semileptonic decays
($P_i(\bar{q_i},q)\rightarrow P_f(\bar{q_f},q)\ell^+\nu_\ell$) has the
structure
\ba\label{QCDformfact}
\left< P_f(p_{f}) \left|\overline{q}_i \gamma_\mu q_f\right|P_i(p_i)\right>
&=&(p_i+p_f)_\mu f_+(q^2)+ (p_i-p_f)_\mu f_-(q^2)\\
&=&\left[(p_i+p_f)_\mu -q_\mu\frac{(m^2_i-m^2_f)}{q^2}\right]f_+(q^2)
+q_\mu\frac{(m^2_i-m^2_f)}{q^2}f_0(q^2),\nonumber
\ea
where $q^\mu$ is the momentum transfer $q^\mu=p^\mu_i-p^\mu_f$. 
In our case $P_f$ is a pion, $P_i$ is a $B$ meson and $q_i=b$. 
For example, to find the $B^0 \rightarrow \pi^-$
formfactors we need then to
evaluate the hadron matrix elements of the quark bilinear $\overline{b}
\gamma_\mu q$, where $q=u$.

Heavy quark and chiral symmetry
transformation properties of chiral currents dictate that the matching of QCD
bilinears onto operators of HMChPT take the form \cite{Falk:1993fr,Wise:1993wa},
\ba
\label{matching}
\bar{b}\gamma^\mu\left(1-\gamma_5\right)q_a &\rightarrow&
i c_L\Tr{\gamma^\mu\left(1-\gamma_5\right)u^\dagger_{ab}H_b(v)},\nonumber \\
\bar{b}\gamma^\mu\left(1+\gamma_5\right)q_a&\rightarrow&
i c_R\Tr{\gamma^\mu\left(1+\gamma_5\right)u_{ab}H_b(v)}.
\ea
The constants $c_L$ and $c_R$ have to be equal because of parity
invariance, therefore we can conclude
\be
\label{vectcurr}
\bar{b}\gamma^\mu q_a\propto\Tr{\gamma^\mu
  \left(u^\dagger_{ab}+u_{ab}\right)H_b(v)}
+\Tr{\gamma_5\gamma^\mu \left(u^\dagger_{ab}-u_{ab}\right)H_b(v)}.
\ee
If no hard pions appear in the final state we can use the
definition of decay constant
\be
\left< 0 \left|\overline{b} \gamma_\mu \gamma_5 q\right|{B}(p_B)\right>
= i F_B p^\mu_B 
\ee
and state $c_L=c_R=\frac{1}{2}F_B \sqrt{m_B}$. Of course this latter result
does not hold for momenta away from $q^2_{max}$  in which case $c_L=c_R$ is just
an effective coupling, as explained in Sect.~\ref{argument}.

In HMChPT it is convenient to use definitions in which the formfactors are
independent of the heavy meson mass
\ba\label{HMChPTformfact}
\left< \pi(p_\pi) \left|\overline{b} \gamma_\mu q\right|B(v)\right>_\mathrm{HMChPT}
=\left[p_{\pi\mu}-\left(v\cdot p_\pi\right)v_\mu\right] f_p(v\cdot p_\pi)
+ v_\mu f_v(v\cdot p_\pi).
\ea
In (\ref{HMChPTformfact}) $v\cdot p_\pi$ is the energy of the pion
in the heavy meson rest frame
\be
v\cdot p_\pi=\frac{m^2_B+m^2_{\pi}-q^2}{2m_B}.
\ee
The formfactors defined in~(\ref{QCDformfact}) and in~(\ref{HMChPTformfact})
are related by matching the relativistic and the HMChPT hadronic current:
\be
\label{relationsff}
f_0(q^2)=\frac{1}{\sqrt{m_B}} f_v(v\cdot p_\pi), \qquad
f_+(q^2)=\frac{\sqrt{m_B}}{2}f_p(v\cdot p_\pi).
\ee
The $\sqrt{m_B}$ factors  in (\ref{relationsff}) are due to the different
normalizations for states used in the two formalisms.
In principle the relations in (\ref{relationsff}) are  valid only
when $q^2\approx q^2_\mathrm{max}$, i.e. when HMChPT is applicable. 
On the other hand, for the arguments shown in 
Sect.~\ref{argument} below, the chiral structure of the formfactors in QCD and
in HMChPT is the same also for $q^2$ away from $q^2_\mathrm{max}$.
However, this does not imply that
(\ref{relationsff}) holds away from $q^2_\mathrm{max}$, at least as far as regard the tree level
term and the leading logarithms.

A matching similar to (\ref{matching}) has to be done also for the relativistic theory described in
Sect.~\ref{relth}. We identify four possible operators\footnote{The last one
is higher order but we included it since it has a different type of contraction
of the Lorentz indices and as an explicit check on the arguments of
Sect.~\ref{argument}.}
\ba
\label{relleftcurr}
J^L_\mu=\frac{1}{2} E_1 t  u^\dagger \nabla_\mu B +  \frac{i}{2} E_2 
t u^\dagger u_\mu B +  \frac{i}{2} E_3 t u^\dagger B^{*}_\mu+ \frac{1}{2}
E_4 t  u^\dagger \left(\nabla_\nu u_\mu \right) B^{* \nu},
\ea
where $E_1$,$\dots$, $E_4$, are effective couplings. $t$ is a constant
spurion vector transforming as $t\rightarrow t g_L^\dagger$,
so that $J^L_\mu$ is invariant under $SU(2)_L$ transformations.
The heavy quark symmetry implies $m_BE_1=E_3$.
Analogously we can introduce a
$J^R_\mu$ current and thus an axial-vector $J^5_\mu=J^R_\mu-J^L_\mu$ and a
vector $J^V_\mu=J^R_\mu+J^L_\mu$ current. They are used respectively to
evaluate the amplitudes of $B\rightarrow \ell \nu_\ell$ and the
$B\rightarrow\pi\ell \nu_\ell$ formfactors as defined in (\ref{QCDformfact}).
We leave the discussion for the latter in Sect.~\ref{ChLogs}, while we quote
here the results of the $B \rightarrow$ vacuum matrix
element at one loop
\ba
\label{decayconst}
F_B=E_1 \left[1+
       \frac{1}{F^2}
       \left(\frac{3}{8}+\frac{9}{8}g^2\right)\overline{A}(m^2_\pi)
\right]
\ea
$\overline{A}(m^2_\pi)$ is defined in (\ref{A}) in the appendix. Here we only quoted
the nonanalytic dependence on the light quark masses for the one-loop part.
%App.~\ref{appendix}. 
We compare (\ref{decayconst}) with the results obtained with HMChPT
\cite{Goity:1992tp}. We
see that $E_1$ plays the role of $F_{H}$ in \cite{Goity:1992tp}
and that the relativistic theory
predicts the same coefficient of the chiral logarithm in $\overline{A}(m^2_\pi)$.

\section{Hard Pion Chiral Perturbation Theory}
\label{argument}

In order to study the chiral behavior of the formfactors at $q^2$ away from
$q^2_\mathrm{max}$ we can not
neglect operators with an arbitrary numbers of derivatives on the external
pion since now its momentum is large. Therefore we must take into account that
the usual power counting of ChPT does not work and we can not be sure a priori
that a loop calculation would make sense.

\begin{figure}
\begin{center}
\includegraphics{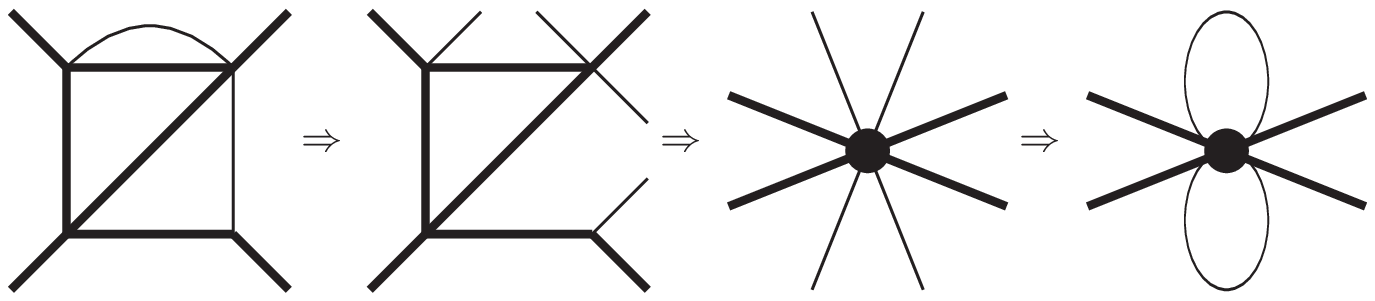}
\end{center}
\caption{An example of the argument used. The thick lines contain a
large momentum, the thin lines a soft momentum. 
Left: a general Feynman diagram with hard and soft lines.
Middle-left: we cut the soft lines to remove the soft singularity.
Middle-right: The contracted version where the hard part is
assumed to be correctly described
by a ``vertex'' of an effective Lagrangian.
Right:
the contracted version as a loop diagram. This is expected to reproduce
the chiral logarithm of the left diagram. Figure from \cite{Bijnens:2009yr}.}
\label{fig:power}
\end{figure}

A similar problem arises in the case of $K_{\ell 3}$ decay. The authors of
\cite{Flynn:2008tg} dealt with it using $SU(2)$ ChPT to study the amplitudes
whether the outgoing pion was soft or not.
They argued they could calculate
the corrections of the type $m_\pi^2 \log{m^2_\pi}$ even in the range of
energy where the usual ChPT does not work. Their argument is based
on the fact that only the soft internal pions are responsible for
the chiral logarithms. These ideas were
generalized by the authors of \cite{Bijnens:2009yr} who made clear
that those arguments basically corresponds to use an effective Lagrangian
to describe the hard part of a general loop calculation in a chiral invariant
way. The situation is shown schematically in Fig.~\ref{fig:power}.
The underlying argument is the same as the analysis for infra-red divergences.
Since the soft lines do not see the hard or short-distance structure of the diagram, we can separate them from the rest of the process. We should thus be able to describe the hard part of any
diagram by an effective Lagrangian. This effective Lagrangian should include
the most general terms allowed consistent with all the symmetries and have
coefficients that depend on the hard kinematical quantities and can even be complex.
A two-loop example will be given in \cite{nextpaper}.
We expect that a proof along the lines of SCET \cite{SCET} should be possible.
Once it is accepted that one can do this, a second step is to
prove that the effective Lagrangian one uses is sufficient to describe
the neighbourhood of the hard process and calculate chiral logarithms.

The latter was done in \cite{Flynn:2008tg} for the case of $K_{\ell3}$
decays by showing that the matrix elements
of operators with higher derivatives was proportional
to the lowest order matrix-element up to terms
of order $m_\pi^2$. In particular, the part including the coefficient
of the chiral logarithms $m_\pi^2\log m_\pi^2$
has the same coefficient relative to the tree level matrix-element
as the lowest order operator. The same was proven for $K\to\pi\pi$
in  \cite{Bijnens:2009yr}.

As a matter of fact, the semileptonic $K$ decay has the same structure
\emph{heavy} $\rightarrow$ \emph{light} as the $B$ one when $M_K$ is treated
as large compared to $m_\pi$ as in \cite{Flynn:2008tg}.
The main
differences between the two processes are the energies involved and that
for the $B$ meson the corresponding vectorial particle $B^*$ is close
to its mass-shell. 
So in order to have a sufficiently complete
effective Lagrangian in the neighbourhood we need to include the $B^*$
as was done in the previous sections.
However, we expect the kind of arguments presented in
\cite{Flynn:2008tg,Bijnens:2009yr} to work here as well.

We note that the effective Lagrangian needs to be complete enough
in the neighbourhood of the underlying process. That also implies
that if we have two different formalisms, both sufficiently complete,
the logarithms should be the same. In particular, the HMChPT
and the relativistic formalism should give the same results.

We are only concerned with terms of order $1$, $m_\pi$ and
$m^2_\pi\log{m^2_\pi}$, i.e. we are not trying to calculate terms of order
$O(m^2_\pi)$ without logarithms.
Here we restrict our discussion to the case
of $SU(2)$ ChPT, and far from $q^2_\mathrm{max}$. It is clear that adding 
(soft) kaon loops does not change the
validity of the arguments. First we analyze the
case of HMChPT and thus we need to look at the matrix elements $\left<
\pi(p_\pi) \left|O\right|B(v)\right>$ where $O$ can be any of the operators in
(\ref{vectcurr}) with possibly more derivatives.
 We want to show that matrix elements of operators with higher number of
derivatives are all proportional to the lowest order ones up to terms
of order $m^2_\pi$  (and without logarithms) which are of higher order.
We need to look at the cases where extra covariant derivatives $D_\mu$ are
added in the operators.
We can distinguish different possibilities depending on which particle
the derivative hits.
\begin{itemize}
\item The case where it hits an internal soft
pion line, leads to $\int d^d p\,p_\mu / (p^2-m^2_\pi)$ which is always
suppressed by three powers of $m_\pi$.
\item If the derivative hits an internal line which is not soft in a loop,
  it is part of the loop diagram that is described by our effective Lagrangian
  and is thus indirectly included via the coefficients. A simple example is
  when a pair of derivatives hits a $B$-meson. 
  Then we get terms like
  $\int d^d p\,p^2_B / (v\cdot p_\pi-\Delta)\dots\simeq m^2_B\int d^d p\,1 /
  (v\cdot p_\pi-\Delta)\dots$, i.e. something that is proportional to the lowest
  order result and that can be included modifying accordingly the effective
  coupling. It corresponds to change the hard structure of the
  loop diagram, what can be described by a proper
  replacement of the effective coupling, as shown in Fig.~\ref{fig:power}.
\item All the extra derivatives should thus act on external lines
  or tree-level internal lines, i.e. those not in a loop. All these
  can thus be transformed into masses of external particles or other
  kinematical quantities, as here $q^2$. None of these has terms of
  order $m_\pi$ or $m_\pi^2\log m_\pi^2$. The kinematical quantities
  we keep fixed and masses have corrections at most of order $m_\pi^2$
  compared to the order $1$ terms and the order $1$ part can be absorbed
  into the coefficient of the lowest order term.
\item Note that if the extra derivatives are contracted with a $v_\mu$ rather
  than another derivative this can also be put into the value of the
  coefficients.
\end{itemize}

Also in the case of the relativistic theory described in Sect.~\ref{relth},
we need to worry if more chiral logarithms arise including operators
like the ones in (\ref{relleftcurr}) but with extra
derivatives. All the above arguments also work except for derivatives that
are contracted with $B^*_\mu$.  
In this case the extra derivatives becomes contracted with the momenta in
the $B^*$ propagator or via $g_{\mu\nu}$ to the external current.
After that the above arguments again apply.
Terms involving a contraction with $B^*$ can always be reduced to the
simplest one which we included in (\ref{relleftcurr}), the $E_4$ term,
and so we have also an explicit test of the last argument.

Near $q^2_\mathrm{max}$ the above arguments fail since kinematical
quantities can contain terms of order $m_\pi$. However, here all pion lines
are soft and we are in the regime of validity of standard HMChPT.

The conclusion from this section is that the coefficient of the chiral
logarithm $m_\pi^2\log m_\pi^2$ is calculable at all values of $q^2$.

\section{The Coefficients of the Chiral Logarithms}
\label{ChLogs}

\begin{figure}
\begin{center}
\includegraphics{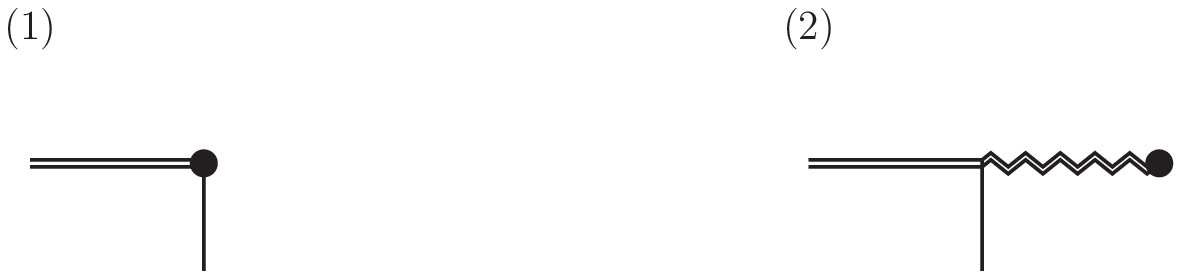}
\end{center}
\caption{The tree-level diagrams contributing to the amplitude. A double line
  correspond to a $B$, a zigzag line to a $B^*$, a single line to a
  pion. A black circle represents the insertion of a $B\rightarrow\pi$ vector
  current.}
\label{fig:treelevel}
\end{figure}

In this section we show the results for the semileptonic decay
$B\rightarrow\pi\ell\nu_\ell$ amplitudes. Hereafter, we quote only the relevant
terms, i.e. the leading ones and the chiral logarithms.
The tree-level diagrams contributing to the amplitude are shown in
Fig.~\ref{fig:treelevel}.
The results for the formfactors at tree level for HMChPT are
\cite{Falk:1993fr,Becirevic:2003ad}
\be
\label{HMChPTtree}
f^\mathrm{Tree}_v(v\cdot p_\pi)=\frac{\alpha}{F},
\qquad
f^\mathrm{Tree}_p(v\cdot p_\pi)=\frac{\alpha}{F}\frac{g}{v\cdot p_\pi+\Delta},
\ee
where $\alpha$ is a constant that takes the value
$\sqrt{m_B/2} F_B$ at $q^2_\mathrm{max}$.
We also have $c_L=c_R=\alpha/\sqrt{2}$.
For the
relativistic theory of Sect.~\ref{relth} we obtain
\be\label{relthtree}
\left.f^\mathrm{Tree}_0(q^2)\right|_{q^2_\mathrm{max}}=\frac{E_1}{F}\frac{1}{4},\qquad \left.f^{\rm
  Tree}_+(q^2)\right|_{q^2_\mathrm{max}}= -\frac{1}{4}\frac{E_3}{F}\frac{m_B}{q^2-m^2_B}g.
\ee
Near $q^2_{max}=(m_B-m_\pi)^2$ the results are obviously the same, since the
propagators in the second equations of (\ref{HMChPTtree}) and
(\ref{relthtree}) become respectively $1/m_\pi$ and $1/(2m_\pi m_B)$. The
different factor of $2$ is due to the different normalization
of states used in HMChPT and in the relativistic formulation.
Note that the relation of the coupling constant to $F_B$ is only valid for
$q^2_{max}=(m_B-m_\pi)^2$. The coupling constant are different at the different
values of $q^2$ and can even be complex. The precise form of (\ref{relthtree})
is only valid near $q^2_\mathrm{max}$. The full expressions are more
complicated.

\begin{figure}
\begin{center}
\includegraphics{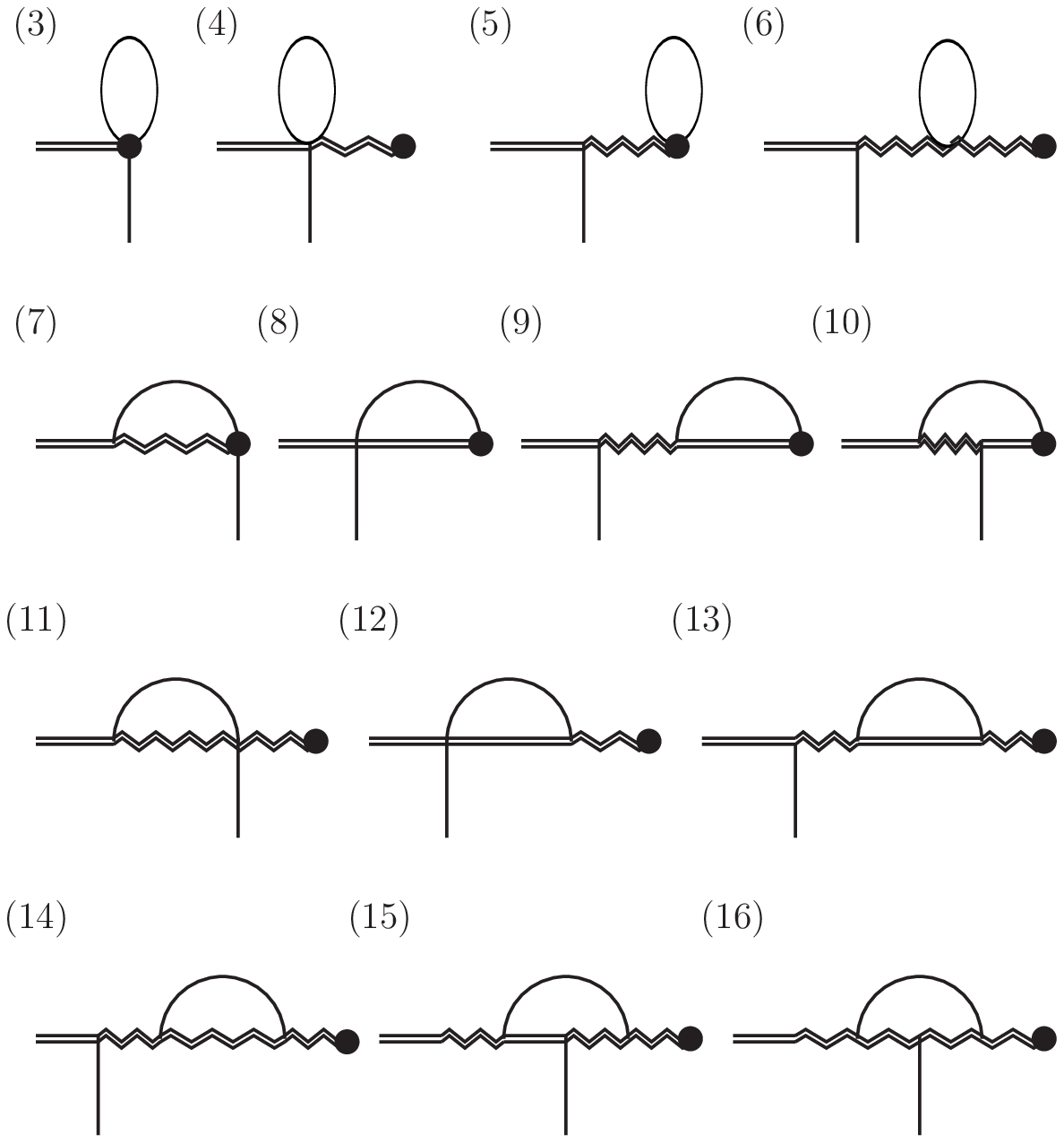}
\end{center}
\caption{The one-loop diagrams contributing to the amplitude.Vertices and
  lines as in Fig.~\ref{fig:treelevel}}
\label{fig:oneloop}
\end{figure}

To proceed with the calculation at one-loop we need the wavefunction
renormalization $Z_\pi$ and $Z_B$. They are the same for HMChPT and the
relativistic theory and read:
\ba
Z_{\pi}=1-\frac{2}{3F^2}\overline{A}(m^2_\pi),
\qquad Z_{B}=1+\frac{9}{4F^2}g^2\overline{A}(m^2_\pi).
\ea
The one-loop diagrams are shown in
Fig.~\ref{fig:oneloop}.
In Tab.~\ref{tab:qsq0} we present the results at $q^2$ away
from $q^2_\mathrm{max}$. To find the results in HMChPT we
expanded the one-loop calculation of  \cite{Becirevic:2003ad}
at $v\cdot p_\pi\rightarrow
m_B$, $m^2_\pi\rightarrow0$. In the relativistic theory we first calculated
the formfactors and then we expanded the loop integrals
for $m^2_\pi \ll m^2_B, (m_B^2-q^2)$. These latter expansions are shown in
App.~\ref{appendix}. We check that the
coefficients of the leading logarithms coincide in the two theories.
Summing up all
the results in Tab.~\ref{tab:qsq0} and including $(1/2) Z_\pi$ and
 $(1/2) Z_B$ times tree level, we find
\ba
\label{finalresult}
f_{v/p}(v\cdot p_\pi)&=&
f^\mathrm{Tree}_{v/p}(v\cdot p_\pi)
\left[1+\left(\frac{3}{8}+\frac{9}{8}g^2\right)\frac{1}{F^2}\overline{A}(m^2_\pi)\right],
\nonumber\\
f_{0/+}(q^2)&=&f^\mathrm{Tree}_{0/+}(q^2)
\left[1+\left(\frac{3}{8}+\frac{9}{8}g^2\right)\frac{1}{F^2}\overline{A}(m^2_\pi)\right],
\ea
i.e. as expected the same coefficients in the two theories. The correction
is also the same for the scalar formfactor $f_0$ and for $f_+$.
\begin{table}
\begin{center}
\begin{tabular}{|c|c|c|}
\hline
Diagram & $f_v$ HMChPT & $f_0$ Rel. Th. \\
\hline
\rule{0cm}{20pt}
\hspace{-0.15cm}(3) & $\frac{5}{24}\frac{1}{F^3}\alpha $&
 $\frac{5}{24F^2}f^\mathrm{Tree}_0(q^2) $\\[4pt]
(8) & $\frac{1}{2}\frac{1}{F^3}\alpha $ &
 $\frac{1}{2F^2}f^\mathrm{Tree}_0(q^2)$
\\[4pt]
\hline
\hline
 & $f_p$ HMChPT & $f_+$ Rel.~Th. \\
\hline
\rule{0cm}{20pt}
\hspace{-0.15cm}(4) & 
 $\frac{2}{6}\frac{g}{F^3}\frac{\alpha}{v\cdot p_\pi + \Delta} $&
 $\frac{1}{3F^2}f^\mathrm{Tree}_+(q^2)$\\[4pt]
(5) & $\frac{3}{8}\frac{ g}{F^3}\frac{\alpha}{v\cdot p_\pi + \Delta} $ &
 $\frac{3}{8F^2}f^\mathrm{Tree}_+(q^2)$
\\[4pt]
\hline
\end{tabular}
\end{center}
\caption{\label{tab:qsq0} The coefficients of the chiral logarithms,
 $\overline{A}(m^2_\pi)$, at
  $q^2$ away from $(m_B-m_\pi)^2$ from the
  different diagrams in Fig.~\ref{fig:oneloop}. The diagrams not listed in the
  table do not contribute with logarithms. The two Lagrangians give the
  same coefficients diagram per diagram provided the tree level coefficients
  are correctly identified.}
\end{table}
\begin{table}
\begin{center}
\begin{tabular}{|c|c|c|}
\hline
Diagram & $f_v$ HMChPT & $f_0$ Rel. Th. \\
\hline
\rule{0cm}{20pt} 
\hspace{-0.15cm}(3) & $\frac{5}{24}\frac{1}{F^3}\alpha $&
 $\frac{5}{96}\frac{1}{F^3}E_1 $\\[4pt]
(8) & $\frac{3}{2}\frac{1}{F^3}\alpha $ &
 $\frac{3}{8}\frac{1}{F^3}E_1 $
\\[4pt]
\hline
\hline
 & $f_p$ HMChPT & $f_+$ Rel. Th. \\
\hline
\rule{0cm}{20pt} 
\hspace{-0.15cm}(4) & $\frac{2}{6}\frac{\alpha }{F^3}\frac{g}{m_\pi}$&
 $\frac{1}{12}\frac{E_3}{F^3}\frac{g}{2m_\pi} $\\[4pt]
(5) & $\frac{3}{8}\frac{\alpha }{F^3}\frac{g}{m_\pi}$ &
 $\frac{3}{32}\frac{E_3}{F^3}\frac{g}{2m_\pi} $
\\[4pt]
(13) & $-\frac{3}{4}\frac{\alpha }{F^3}\frac{g^3}{m_\pi}$ &
 $\frac{1}{16}\frac{E_3}{F^3}\frac{g^3}{2m_\pi} $
\\[4pt]
(14) & $\frac{3}{2}\frac{\alpha }{F^3}\frac{g^3}{m_\pi}$ &
 $\frac{1}{8}\frac{E_3}{F^3}\frac{g^3}{2m_\pi} $
\\[4pt]
(15) & $\frac{1}{12}\frac{\alpha }{F^3}\frac{g^3}{m_\pi}$ &
 $\frac{1}{48}\frac{E_3}{F^3}\frac{g^3}{2m_\pi} $
\\[4pt]
(16) & $-\frac{1}{6}\frac{\alpha}{F^3}\frac{g^3}{m_\pi}$ &
 $-\frac{1}{24}\frac{E_3}{F^3}\frac{g^3}{2m_\pi} $
\\[4pt]
\hline
\end{tabular}
\end{center}
\caption{\label{tab:qsqmax} The coefficients of the chiral logarithms,
  $\overline{A}(m^2_\pi)$, at
  $q^2_\mathrm{max}$ from the
  different diagrams in Fig.~\ref{fig:oneloop}. The  diagrams not listed in
  the table do not contribute to the chiral logarithms. 
  The two Lagrangians give the
  same coefficients provided the tree level coefficients are correctly
  identified.}
\end{table}
In  Tab.~\ref{tab:qsqmax} we quote also the
results in the limit $q^2=q^2_{\max}=(m_B-m_\pi)^2$ where the two theories
must give the same outcome, being one the relativistic limit of the
other. So this is another check of the validity of our relativistic theory.
 Summing up all
the results as explained above we find at $q^2_\mathrm{max}$
\ba
\label{finalresult2}
f_{v}(v\cdot p_\pi)&=&f^\mathrm{Tree}_{v}(v\cdot p_\pi)\left[1+\left(\frac{11}{8}+\frac{9}{8}g^2\right)\frac{1}{F^2}\overline{A}(m^2_\pi)\right],\nonumber\\
f_{0}(q^2)&=&f^{\rm
  Tree}_{0}(q^2)\left[1+\left(\frac{11}{8}+\frac{9}{8}g^2\right)\frac{1}{F^2}\overline{A}(m^2_\pi)\right],\nonumber\\
f_{p}(v\cdot p_\pi)&=&f^\mathrm{Tree}_{p}(v\cdot p_\pi)\left[1+\left(\frac{3}{8}+\frac{43}{24}g^2\right)\frac{1}{F^2}\overline{A}(m^2_\pi)\right],\nonumber\\
f_{+}(q^2)&=&f^\mathrm{Tree}_{+}(q^2)\left[1+\left(\frac{3}{8}+\frac{43}{24}g^2\right)\frac{1}{F^2}\overline{A}(m^2_\pi)\right],
\ea
i.e. agreement between the two theories.

As a final check, we notice that the results obtained including
only those diagrams where no $B^*$ appears (i.e. (1),(3) and (8) in
Fig.~\ref{fig:treelevel} and \ref{fig:oneloop}) coincide with the ones
in~\cite{Flynn:2008tg} for the
$K\rightarrow\pi$ amplitudes. The chiral corrections must agree in these two
cases since, as we remarked above, the only difference between the two
processes is the presence of the vectorial $B^*$ particle.

\section{Conclusions}

In this paper we have calculated the pionic logarithms in the semileptonic
$B\rightarrow\pi$ and $D\rightarrow\pi$ transitions. We have reproduced the
known results near the endpoint $q^2=(m_B-m_\pi)^2$, Eq.~(\ref{finalresult2})
and obtained the chiral logarithm also away from the endpoint
in Eq.~(\ref{finalresult}) and it was the same for both formfactors.

\section*{Acknowledgments}

IJ gratefully acknowledges an Early Stage Researcher position supported by the
EU-RTN Programme, Contract No. MRTN--CT-2006-035482, (Flavianet)
This work is supported in part by the European Commission RTN network,
Contract MRTN-CT-2006-035482  (FLAVIAnet), European Community-Research
Infrastructure Integrating Activity ``Study of Strongly Interacting Matter'' 
(HadronPhysics2, Grant Agreement n. 227431)
and the Swedish Research Council. This work heavily used FORM
\cite{Vermaseren:2000nd}.

\appendix

\section{Loop integrals expansions}
\label{appendix}

We collect the relevant expansions of the one-loop integrals needed to evaluate
the diagrams in Fig.~\ref{fig:oneloop} in the framework of the relativistic
theory of Sect.~\ref{relth}. In the calculation we need the
one-, two- and three-point
functions defined as ($d=4-2\epsilon$)
\ba
\label{A}
A(m^2_1)&=&\frac{1}{i}\int{\frac{d^dk}{(2\pi)^d}\frac{1}{k^2-m^2_1}}\\
\label{B}
B(m^2_1,m^2_2,p^2)&=&
\frac{1}{i}\int{\frac{d^dk}{(2\pi)^d}\frac{1}{(k^2-m^2_1)((p-k)^2-m^2_2)}}\\
\label{C}
C(m^2_1,m^2_2,m^2_3,p^2_1,p^2_2,q^2)&=&\frac{1}{i}
\int{\frac{d^dk}{(2\pi)^d}
\frac{1}{(k^2-m^2_1)((k-p_1)^2-m^2_2)((k-p_1-p_2)^2-m^2_3)}},\nonumber\\
\ea
with $q^2 =(p_1+p_2)^2$.
Actually two- and three-point functions with extra powers of momenta in the
numerator contribute too, but we do not intend to give their
definitions here. They can be found in 
\cite{Bijnens:2002hp} in precisely the form used here.
We only stress that all these
functions can be rewritten in terms of (\ref{A}), (\ref{B}) and (\ref{C})
\cite{Passarino:1978jh}. 
The finite parts of $A(m^2_1)$ and $B(m^2_1,m^2_2,q^2)$
are \cite{'tHooft:1978xw}
\ba
\label{Abar}
\overline{A}(m^2_1)&=&-\frac{m^2_1}{16\pi^2}\log{\left(\frac{m^2_1}{\mu^2}\right)},\\
\label{Bbar}
\bar{B}(m^2_1,m^2_2,q^2)&=&
\frac{1}{16\pi^2}\left[-1-\int^1_0dx
\log{\left(\frac{m_1x+m_2(1-x)-x(1-x)q^2}{\mu^2}\right)}\right].
\ea
In the calculation of the amplitude the three-point function
$C(m^2_1,m^2_2,m^2_3,p^2_1,p^2_2)$
always depends on the masses as $(m^2,M^2,M^2,m^2,q^2)$ with $m=m_\pi$,
and $M=m_B$. It can be
rewritten using Feynman parameters $x$, $y$
\ba
\label{Cbar}
C(m^2,M^2,M^2,M^2,m^2,q^2)&=&-\frac{1}{16\pi^2}\int^1_0dx\int^{1-x}_0\hspace{-0.4cm}dy\left[m^2(1-x-2y+y^2)+M^2(x+y)^2\right.
\nonumber\\&&\left.+(q^2-M^2-m^2)(-y+y(x+y))\right]^{-1}.
\ea
In order to find the appropriate chiral logarithms we expanded (\ref{Bbar}) and
(\ref{Cbar}) for small $m^2/M^2$ . We quote only the terms of the
expansions containing the chiral logarithms $\log{(m^2/\mu^2)}$
\ba  
\bar{B}(m^2,M^2,q^2)& =& -\frac{1}{M^2-q^2}\overline{A}(m^2),\qquad q^2\ll q^2_\mathrm{max}\\
C(m^2,M^2,M^2,m^2,q^2)&=&-\frac{1}{(M^2-q^2)^2}\overline{A}(m^2),\qquad q^2\ll q^2_\mathrm{max}\\
\bar{B}(m^2,M^2,M^2)& = &\frac{1}{2M^2}\overline{A}(m^2),\\
\bar{B}(m^2,M^2,m^2) &=& 0,\\
\bar{B}(m^2,M^2,(M-m)^2)&=&-\frac{1}{mM}\overline{A}(m^2)-\frac{1}{M^2}\overline{A}(m^2).
\ea
The expansions of the three-point functions at $q^2_\mathrm{max}$ are a bit more
involved. The reason is that the reduction formulas present a singularity at
$q^2_\mathrm{max}=(M-m)^2$ for $m^2=0$.
Thus we expand each of them directly, from the Feynman parameter integral,
without rewriting them in terms of (\ref{A}), (\ref{B}) and (\ref{C}).
To do this one rewrites the integral in (\ref{Cbar}) using $z=x+y$
as
\ba
C(m^2,M^2,M^2,m^2,(M-m)^2)&=&
-\frac{1}{16\pi^2}\int^1_0dz\int^{z}_0dy \times
\nonumber\\&&
\hspace*{-2cm}\frac{1}{\left[M^2 z^2+m^2 +2mMy+\left(m^2(-z-y+y^2)-2mMyz\right)\right]}.
\ea
The part in the denominator in brackets is always suppressed by at least $m/M$
compared to the first three terms for all values of $z$ and $y$ and we can
thus expand in it. The remaining integrals can be done with elementary means.
The result of the expansion is, quoting only up to the order needed for
this work,
\ba
C(m^2,M^2,M^2,M^2,m^2,(M-m)^2 )&=&
-\frac{1}{2}\left(\frac{1}{m^2M^2}\overline{A}(m^2)
       +\frac{1}{mM^3}\overline{A}(m^2)+\frac{1}{M^4}\overline{A}(m^2)\right),
\nonumber\\
\\
\overline{C}_{11}(m^2,M^2,M^2,M^2,m^2,(M-m)^2 ) &=&
\frac{1}{2}\frac{1}{mM^3}\overline{A}(m^2)+ \frac{7}{12}\frac{1}{M^4}\overline{A}(m^2),
\\
\overline{C}_{12}(m^2,M^2,M^2,M^2,m^2,(M-m)^2 ) &=&
\frac{1}{3}\frac{1}{mM^3}\overline{A}(m^2)+ \frac{7}{12}\frac{1}{M^4} \overline{A}(m^2),
\\
\overline{C}_{21}(m^2,M^2,M^2,M^2,m^2,(M-m)^2 ) &=&
 -\frac{1}{6}\frac{1}{M^4}\overline{A}(m^2),
\\
\overline{C}_{22}(m^2,M^2,M^2,M^2,m^2,(M-m)^2 ) &=& -\frac{7}{30}\frac{1}{M^4}\overline{A}(m^2),\\
\overline{C}_{23}(m^2,M^2,M^2,M^2,M^2,m^2,(M-m)^2 )&=&-\frac{1}{4}\frac{1}{M^4}\overline{A}(m^2),\\
\overline{C}_{24}(m^2,M^2,M^2,M^2,m^2,(M-m)^2 )&=&-\frac{1}{12}\frac{1}{M^2}\overline{A}(m^2).
\ea
The other three-point functions do not give any leading logarithm.

\end{document}